# Coherent Topological Transport on the Surface of Bi$_2$Se$_3$

Dohun Kim[1], Paul Syers[1], Nicholas P. Butch[2], Johnpierre Paglione[1], and Michael S. Fuhrer[1*]

1. Center for Nanophysics and Advanced Materials, Department of Physics, University of Maryland, College Park, MD 20742-4111, USA

2. Condensed Matter and Materials Division, Lawrence Livermore National Laboratory, Livermore, CA 94550, U.S.A.

**The two-dimensional (2D) surface state of the three-dimensional strong topological insulator (STI) is fundamentally distinct from other 2D electron systems[1] in that the Fermi arc encircles an odd number of Dirac points[2-8]. The TI surface is in the symplectic universality class and uniquely among 2D systems remains metallic and cannot be localized by (time-reversal symmetric) disorder[9]. However, in finite-size samples inter-surface coupling can destroy the topological protection[10-12]. The question arises: At what size can a thin TI sample be treated as having decoupled topological surface states? We show that weak anti-localization[13-21] (WAL) is extraordinarily sensitive to sub-meV coupling between top and bottom topological surfaces, and the surfaces of a TI film may be coherently coupled even for thicknesses as large as 12 nm. For thicker films we observe the signature of a true 2D topological metal: perfect weak anti-localization in quantitative agreement with two decoupled surfaces in the symplectic symmetry class.**

Weak localization (WL) and weak anti-localization (WAL) describe corrections to the classical electrical conductivity of two-dimensional electron gases due to coherent interference of time-reversed paths. The TI surface state is in the symplectic symmetry class and expected to exhibit perfect WAL described by Higami, Larkin and Nagaoka[13]:

$$\Delta\sigma(H) = \alpha \frac{e^2}{\pi h}[\ln\frac{H_0}{H} - \psi\left(\frac{1}{2} + \frac{H_0}{H}\right)] \qquad (1)$$

where $H_0 = \hbar/4De\tau_\varphi$ is a characteristic field related to the phase coherence time $\tau_\varphi$ and diffusion constant $D$, $\psi$ is the digamma function, and $\alpha$ is the overall amplitude whose expected value for a single 2D channel is 1/2. However, WAL is sensitive to the competition between the phase coherence time $\tau_\varphi$ and other time scales[22]; in TI thin films, carriers may scatter into additional conducting channels (from top to bottom surface, or into the conducting bulk), modifying the WAL behavior. In contrast to previous studies of WAL in TI thin films[13-21] where strong surface to bulk scattering dominates inter-channel coupling, we measure WAL in gate-tuned, bulk insulating $Bi_2Se_3$ thin films where we expect negligible surface-bulk scattering. We find that the WAL behavior is governed by the ratio of $\tau_\varphi$ to the inter-surface tunneling time $\tau_t = h/2\Delta$ where $h$ is Planck's constant and $\Delta$ is the hybridization gap induced by inter-surface tunneling[10,12]. For thick films, $\tau_\varphi/\tau_t < 1$ and we observe WAL according to Eqn. [1] with $\alpha = ½ + ½ = 1$ corresponding to two decoupled TI surfaces each with $\alpha = ½$. A thickness and doping-dependent crossover is observed when $\tau_\varphi/\tau_t > 1$ to a regime where $\alpha = ½$, indicating the coherent coupling of two TI surfaces. For even thinner samples, $\Delta$ becomes comparable to the disorder strength and the Berry's phase is completely randomized at low carrier density causing the suppression of WAL and WL, reflecting the loss of topological protection for strongly coupled surfaces.

We study mechanically exfoliated $Bi_2Se_3$ single crystals[23] ranging in thickness from 5 to 17 quintuple layers (QL). Figure 1a shows a representative device (thickness 12 QL), where height profiles across the width direction of the 17, 12, 7, and 5 QL devices are shown in figure 1b. In order to achieve the topological transport regime we employed molecular charge transfer $p$-type doping by thermal evaporation of 2,3,5,6-tetrafluoro-7,7,8,8-tetracyanoquinodimethane, described previously[24] (see also Methods and Supplementary Information S1).

Figure 1c and d show the longitudinal resistivity $\rho_{xx}$ and Hall carrier density $n_H = 1/(eR_H)$ (where $R_H$ is the Hall coefficient, and $e$ is the elementary charge) of the devices at a temperature of 2K as a function of back gate voltage $V_g$. We measure $n_H$ in the range of 2-7 x $10^{12}$ cm$^{-2}$ at zero gate voltage. The carrier density in the topological surface state at the bulk conduction band edge is approximately 1.0 x $10^{13}$ cm$^{-2}$ hence the devices are in the TI regime before application of a back gate voltage. Ambipolar electric field effects are indicated by the sharp peak of $\rho_{xx}(V_g)$ and the sign change in $n_H$ at the charge neutrality points ($V_{g,0}$, figure. 1c dashed lines). The dependence of $\rho_{xx}(V_{g,0})$ on temperature $T$ in all of our devices, including 5 QL, shows metallic behavior (See Supplementary Information S2) which is likely due to conduction through inhomogeneity-driven electron-hole puddles[24,25]. However the thinnest (5 QL) device shows an anomalously large maximum $\rho_{xx}(V_{g,0})$ of about 23 kΩ which cannot be understood within the self-consistent theory for a Dirac band in the presence of charge disorder[24,26,27], suggesting that the inter-surface hybridization gap Δ becomes important in determining the resistivity in this regime[10]. We note that true insulating behavior (divergent $\rho_{xx}$ as $T \to 0$) is not observed for the 5 QL sample, but was previously observed for 3QL $Bi_2Se_3$ [11]. More work is needed to understand the $\rho_{xx}$ maximum in the 5 QL sample (See Supplementary Information S2 for further discussion).

We now turn to discuss gate-tuned WAL behavior in the TI regime. Figure 2 shows the magneto-conductivity $\Delta\sigma(H)$ for all four devices. Curves are taken at similar carrier density $n = C_g(V_g - V_{g,0})/e$ (where gate capacitance $C_g \approx 11$ nFcm$^{-2}$) ranging from $\approx 7 \times 10^{12}$ cm$^{-2}$ ($n$-type) to $-2 \times 10^{12}$ cm$^{-2}$ ($p$-type) except for the 17 QL device where only $n$-type carrier density (7 and 2 $\times 10^{12}$ cm$^{-2}$) could be observed due to relatively high initial doping. The data for the entire range of $n$ and thickness can be fitted (dashed curves in Fig. 2) well with Eqn. (1). The validity of Eqn. (1) in the multichannel limit has been addressed in a number of previous studies[14, 18, 19, 28], finding that Eqn. [1], in particular the logarithmic correction term, provides a robust physical description of WAL behavior even in the multichannel limit irrespective of the Drude conductivity of each channel[19, 28]. Therefore $\alpha = m/2$ probes the number of channels $m$.

Figure 3 shows the variation of α obtained from the fit to Eqn. (1) as a function of $n$ for 17 (black), 12 (red), 7 (green), and 5 (blue) QL devices measured at 2 K. In all devices α is close to 1/2 at high $n \approx 8 \times 10^{12}$ cm$^{-2}$, which we interpret as WAL in a single strongly coupled coherent channel[14-16, 18, 19]. However, the behavior of α upon gate tuning shows very different behaviors depending on the thickness of the devices. The thickest device (17 QL) shows WAL consistent with two decoupled top and bottom topological surfaces (α ≈ 1) starting at moderate $n \approx 6 \times 10^{12}$ cm$^{-2}$, maintained down to the minimum accessible $n$. Similar variation of α ranging from 0.7 to 1 was observed in Ref. [19] with the application of top gate, and interpreted there as the decoupling of top surface and bulk (plus bottom) channels due to formation of band-bending induced depletion layer. We do not rule out the possible contribution of bulk-surface scattering in 17 QL at high $n$. However, we note that this effect alone cannot explain the general behavior of α for low $n$ and thinner devices. Notably, we identify two crossovers in the 12 QL device: we observe sharp transition of α from ≈ 1/2 to ≈ 1 near $n \approx 1 \times 10^{12}$ cm$^{-2}$ ($n$-type), and back to 1/2 at

$n \approx -1 \times 10^{12}$ cm$^{-2}$ (*p*-type). $\Delta\sigma(H)$ in our devices include a moderate contribution from (universal) conductance fluctuations (Fig. 2), as commonly observed in micro-fabricated Bi$_2$Se$_3$ devices[15,20]. However we repeated similar WAL measurements at 2 K for the 12 QL device on five different thermal runs, where the conductance fluctuation contribution is randomized, and find that the crossovers are reproducible within the experimental uncertainty (Fig. 3, error bars). This transition is absent in the 7 QL device where WAL in the entire range of *n* indicates a strongly coupled single channel ($\alpha \approx 1/2$). Finally, at an even smaller thickness (5QL), we observe strong suppression of WAL for $-1 \times 10^{12}$ cm$^{-2}$ < $n$ < $1 \times 10^{12}$ cm$^{-2}$, (inset of Fig. 3, see also Fig. 2d).

We now examine the coherence time $\tau_\varphi$ and compare to the estimated interlayer tunnel time $\tau_t$. We estimate $\tau_\varphi = \hbar/4eDH_0$ from the fits to Eqn. [1] and using $D = \tau v_f^2 /2$ where Fermi velocity $v_f \approx 3 \times 10^7$ cm/s for Bi$_2$Se$_3$ [29], and the momentum relaxation time $\tau$ is calculated from the measured $\sigma(n)$. $\tau_t$ represents the characteristic time of transition (half the period Rabi oscillations) between localized states in two quantum wells with energy splitting $\Delta$, thus $\tau_t = h/2\Delta$. We estimate $\Delta$ in Bi$_2$Se$_3$ by fitting the existing experimental data[10] to an exponentially decaying function, $\Delta(t) = [992 \text{ meV}]e^{-0.67[t \text{ (nm)}]}$, and obtain $\Delta$ of 34.8, 9.1, 0.3, and 0.01 meV for 5, 7, 12, and 17 QL, respectively. (Theoretical calculations of finite size effects in Bi$_2$Se$_3$ predict systematically smaller $\Delta$ at a given thickness and also suggest that $\Delta$ shows oscillatory decaying behavior[12], which do not seem to be observed in the previous work[10] or this work.)

Figure 4 shows the estimated $\tau_\varphi$ and $\tau_t$ as functions of *n* for the different thickness samples. Upon gate tuning, $\tau_\varphi$ changes by an order of magnitude ranging from $\approx 4$ to $\approx 50$ *ps*, and shows a sharp dip near $n = 0$. For the 17 QL device, $\tau_\varphi/\tau_t \ll 1$ in the entire range of *n*. In this limit electrons on either surface lose coherence before scattering to the other, thus each surface acts as an independent coherent transport channel and $\alpha \approx 1$. In contrast a crossover occurs from $\tau_\varphi/\tau_t \gg 1$ at high unipolar n- and p-type carrier densities to $\tau_\varphi/\tau_t < 1$ near $n = 0$ in the 12 QL device, which is consistent with the observed crossover from $\alpha = \frac{1}{2}$ to $\alpha = 1$ (Fig. 3). In the thinner device (7 QL), the condition $\tau_\varphi/\tau_t \gg 1$ is satisfied, and the two surfaces are strongly coupled ($\alpha \approx 1/2$) for the entire range of n. In contrast to semi-classical Boltzmann transport in $Bi_2Se_3$ TI films, where the signature of finite size effect appear only in the ultrathin limit ($\leq$ 3QL)[11], phase coherent transport offers an exquisitely sensitive probe of the hybridization of top and bottom transport channels, detecting a gap $\Delta$ as small as 0.3 meV in the 12 QL TI film[10].

The behavior described above is in general consistent in the 5 QL device except near the Dirac point, where WAL is strongly suppressed. The suppression of WAL can be understood in terms of the expected change of Berry phase $\theta_B = \pi(1 - \Delta/2E_f)$ in TIs[30] as a function of $\Delta$ and the Fermi energy $E_f$ ; $\theta_B$ is reduced from $\pi$ when hybridization induced gap opens. Assuming a gapped Dirac dispersion $E_f = \sqrt{(\hbar v_f k_f)^2 + (\Delta/2)^2}$, we estimate $E_f \leq 52$ meV and $\theta_B \leq 0.67\pi$ at $|n| \leq 1 \times 10^{12}$ cm$^{-2}$. This range of n is comparable to the electron-hole puddle density $n^* \approx 0.5 \times 10^{12}$ cm$^{-2}$ per surface (Fig. 1d) hence transport occurs through a landscape of electron and hole puddles with $\theta_B$ spanning the entire range $0 \leq \theta_B \leq \pi$. It appears that in this regime of highly inhomogeneous Berry phase that both WAL and WL (expected as $\theta_B \to 0$) are suppressed. We note that similar suppression of WAL was recently observed in thin epitaxial $Bi_2Se_3$ of varying thickness (and $\Delta$) but fixed $E_f$ [21], and a competition between WAL and WL was observed in gated 4 QL epitaxial $(Bi_{0.57}Sb_{0.43})_2Te_3$ [31]. Our observation of the WAL suppression by tuning both $E_f$ and $\Delta$ (thickness) allows us to identify the ratio of $\tau_\varphi/\tau_t$ as the driver for the crossover between coupled and decoupled surfaces, and the ratio of $\Delta$ to disorder strength as the driver for the crossover to the regime of suppressed WAL/WL.

**Methods**

Low-doped (carrier density ~ $10^{17}$ cm$^{-3}$) bulk Bi$_2$Se$_3$ single crystals with bulk resistivity exceeding 2 m$\Omega$ cm at 300 K were grown by melting high purity bismuth (6N) and selenium (5N) in sealed quartz ampoules[23]. Crystals were exfoliated with Scotch tape and deposited on doped Si covered with 300nm SiO$_2$. Thin Bi$_2$Se$_3$ crystals with thickness ranging from 5 to 17 nm were identified by combined use of optical and atomic force microscopy (AFM). From the AFM height profile of Bi$_2$Se$_3$ thin flakes, atomically flat surfaces with height variation less than 1 QL were chosen as channel area. Electron beam lithography, thermal evaporation and liftoff techniques were used to make electrical contact (Cr/Au: 5/70 nm). For accurate determination of the geometric factor, thin films were patterned into Hall bar geometry (see Fig. 1a) with typical aspect ratio ($L/W$) of about 2 and shortest length exceeding 2 µm using Ar plasma at a pressure of ~6.7 Pa (5 x 10$^{-2}$ Torr). Molecular charge transfer doping was done by thermal evaporation of ~10 nm of 2,3,5,6-tetrafluoro-7,7,8,8-tetracyanoquinodimethane organic molecules (Aldrich) on top of the fabricated samples[24].

Transport measurements were performed using standard four-probe ac methods with low frequency (<17 Hz) excitation currents (rms amplitude 100 nA) using Stanford Research Systems SR830 Lock in amplifiers and a commercial cryostat equipped with 9 T superconducting magnet. For the 12 QL device, five different WAL measurements as a function of gate voltage at 2 K were conducted, where the sample was warmed up to 300 K and exposed to air for a few hours between thermal runs.

Acknowledgements

This work was supported by NSF grant number DMR-1105224. Preparation of $Bi_2Se_3$ was supported by NSF MRSEC (DMR-0520471) and DARPA-MTO award (N66001-09-c-2067). NPB was partially supported by the Center for Nanophysics and Advanced Materials. D. Kim acknowledges useful conversations with Sergey S. Pershoguba.

Author Contributions

D.K. fabricated devices, performed the electrical measurements, and analyzed the data with M.S.F. P.S., N.P.B, and J.P. prepared single crystal $Bi_2Se_3$ starting material. D.K. and M.S.F. wrote the manuscript with contributions from all authors.

Additional Information

The authors declare no competing financial interests. Supplementary information accompanies this paper. Correspondence and requests for materials should be addressed to Michael S. Fuhrer (mfuhrer@umd.edu).

**Figure captions**

**Figure 1 Characterization of Bi$_2$Se$_3$ Hall bar devices. a,** AFM image of 12 QL Bi$_2$Se$_3$ Hall bar device. **b,** Height profiles along width directions of the devices used in this study. **c,** Longitudinal resistivity $\rho_{xx}$ and **d**, sheet carrier density $n_H$ determined from Hall measurement as a function of back gate voltage at the temperature of 2 K for 17 (black), 12 (red), 7 (green), and 5 (blue) QL devices. The dashed lines show charge neutrality points.

**Figure 2 Weak anti-localization in the topological insulator regime.** Magneto-conductivity $\Delta\sigma$ as a function of perpendicular magnetic field $H$ in **a,** 17, **b,** 12, **c,** 7, and **d,** 5QL devices measured at 2 K at gate-induced carrier densities indicated in the legends. Dashed curves show least-square fits to Eqn. [1]. Zeros of all curves are offset by 0.7 $e^2/h$ for clarity.

**Figure 3 Coupled / decoupled coherent transport in Bi$_2$Se$_3$ surface states.** Variation of the amplitude of weak anti-localization $\alpha$ as a function of 2D carrier density $n$ for 17 (black square), 12 (red circle), 7 (green triangle), and 5 QL (blue triangle) thick devices measured at 2K. For the 12 QL device, the error bars show standard deviations determined from WAL measurements taken at five different thermal runs. The inset shows detailed behavior of suppression of WAL for 5 QL at small $n$.

**Figure 4 Phase coherence time ($\tau_\varphi$) vs. interlayer tunneling time ($\tau_t$).** Comparison of phase coherence time $\tau_\varphi$ determined from fit to Eqn. [1] and transport scattering time to interlayer tunneling time $\tau_t$ estimated from surface hybridization induced energy gap $\Delta$ as a function of 2D

carrier density $n$. Symbols are experimentally measured $\tau_\varphi$ for 17 (black square), 12 (red circle), 7 (green triangle), and 5 QL (blue triangle) devices. The dashed lines with corresponding colors show estimated inter-surface tunneling time $\tau_t$.

Figure 1.

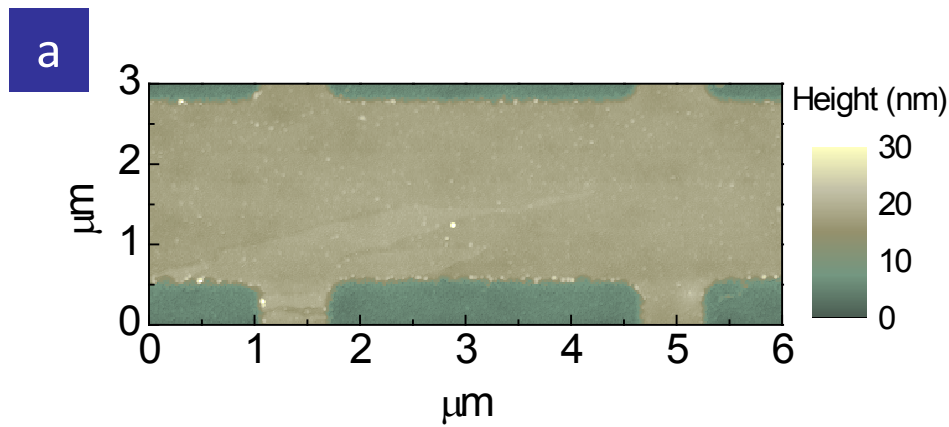
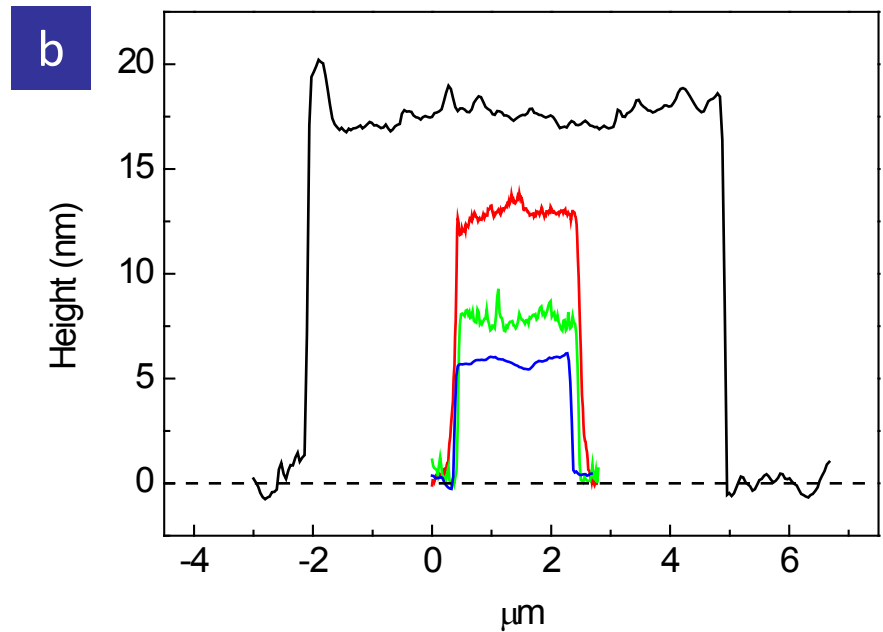
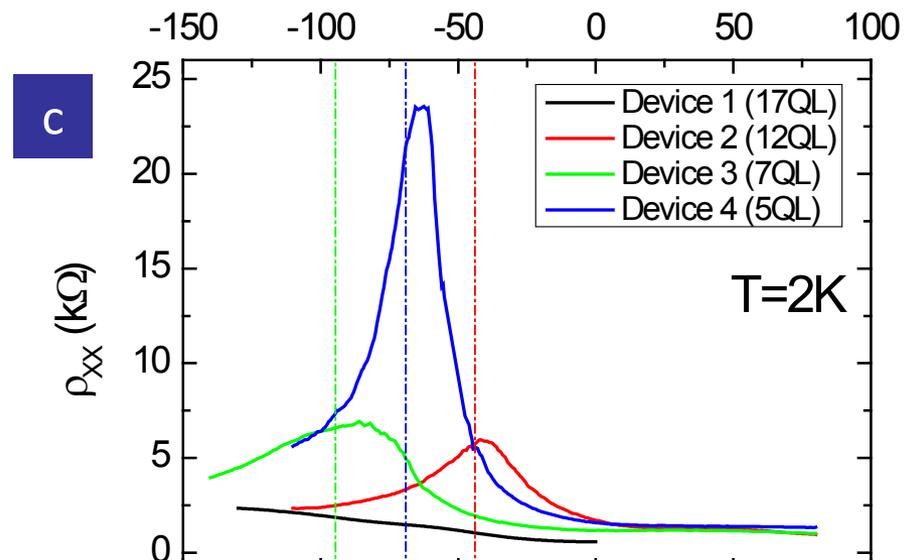
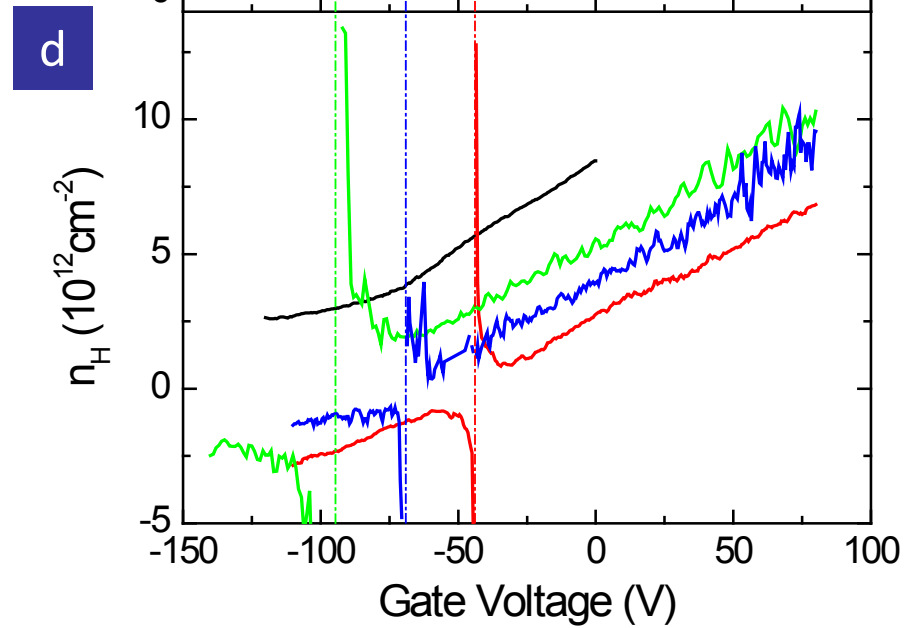

Figure 2.

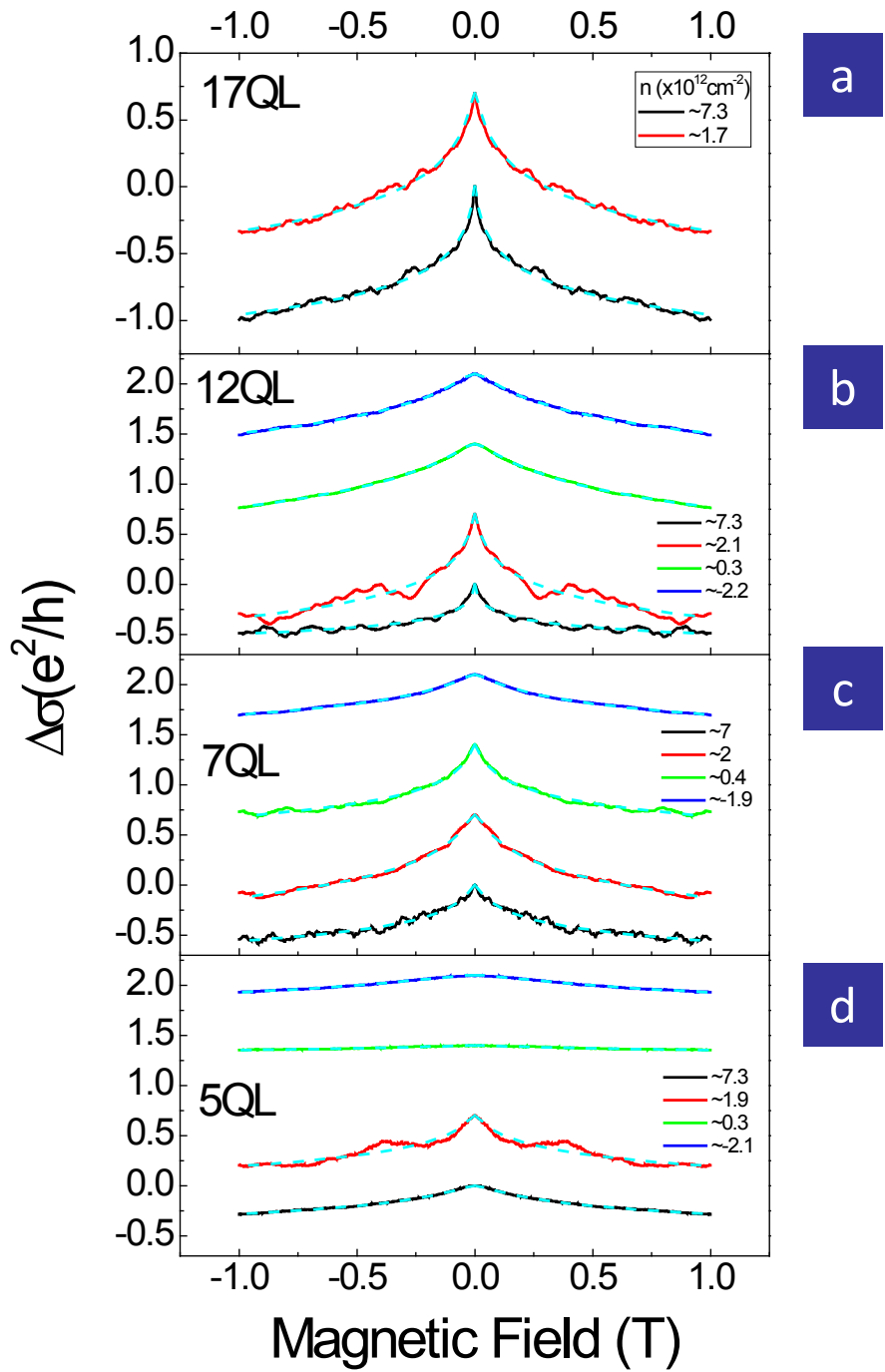

Figure 3.

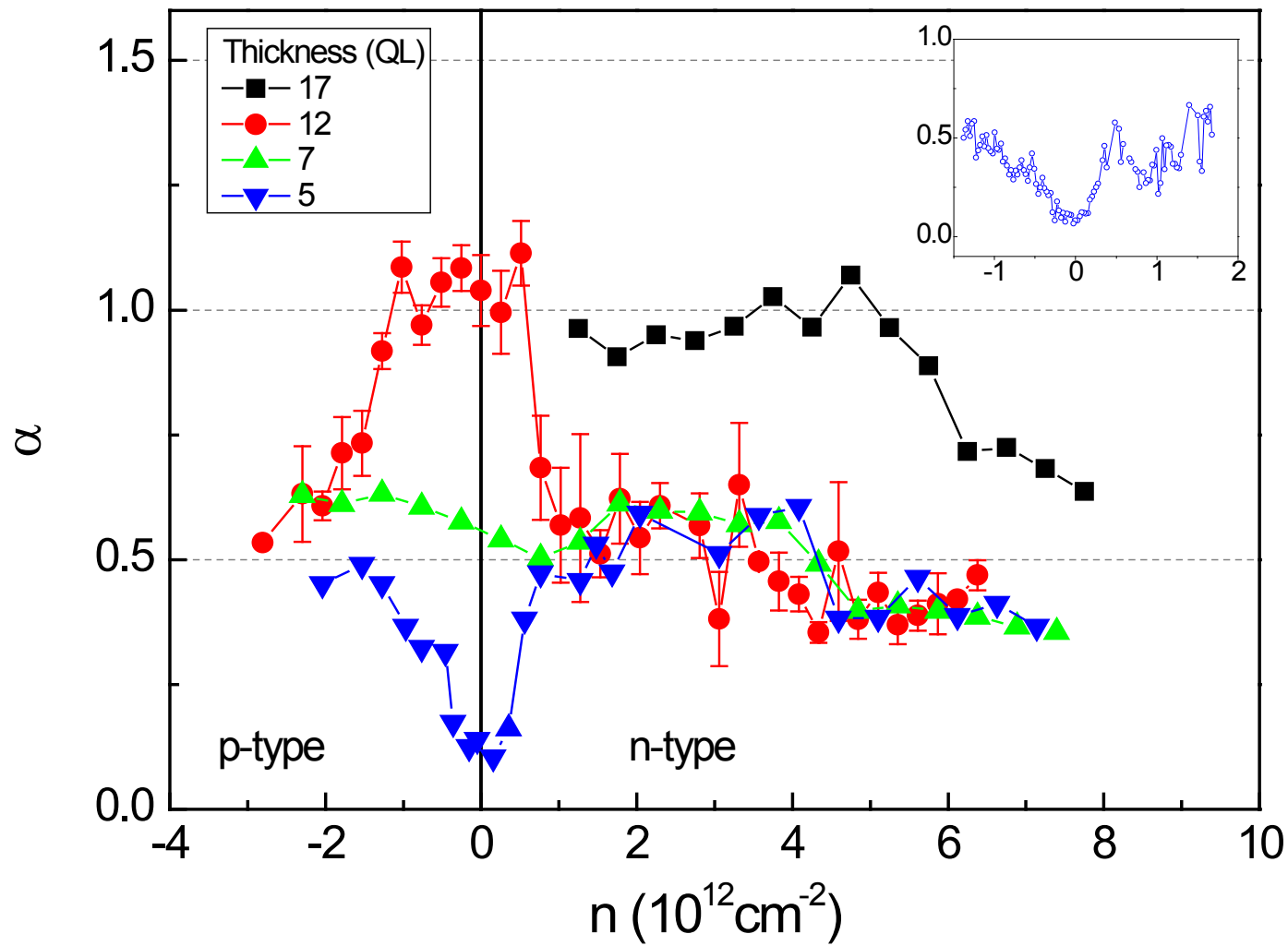

Figure 4.

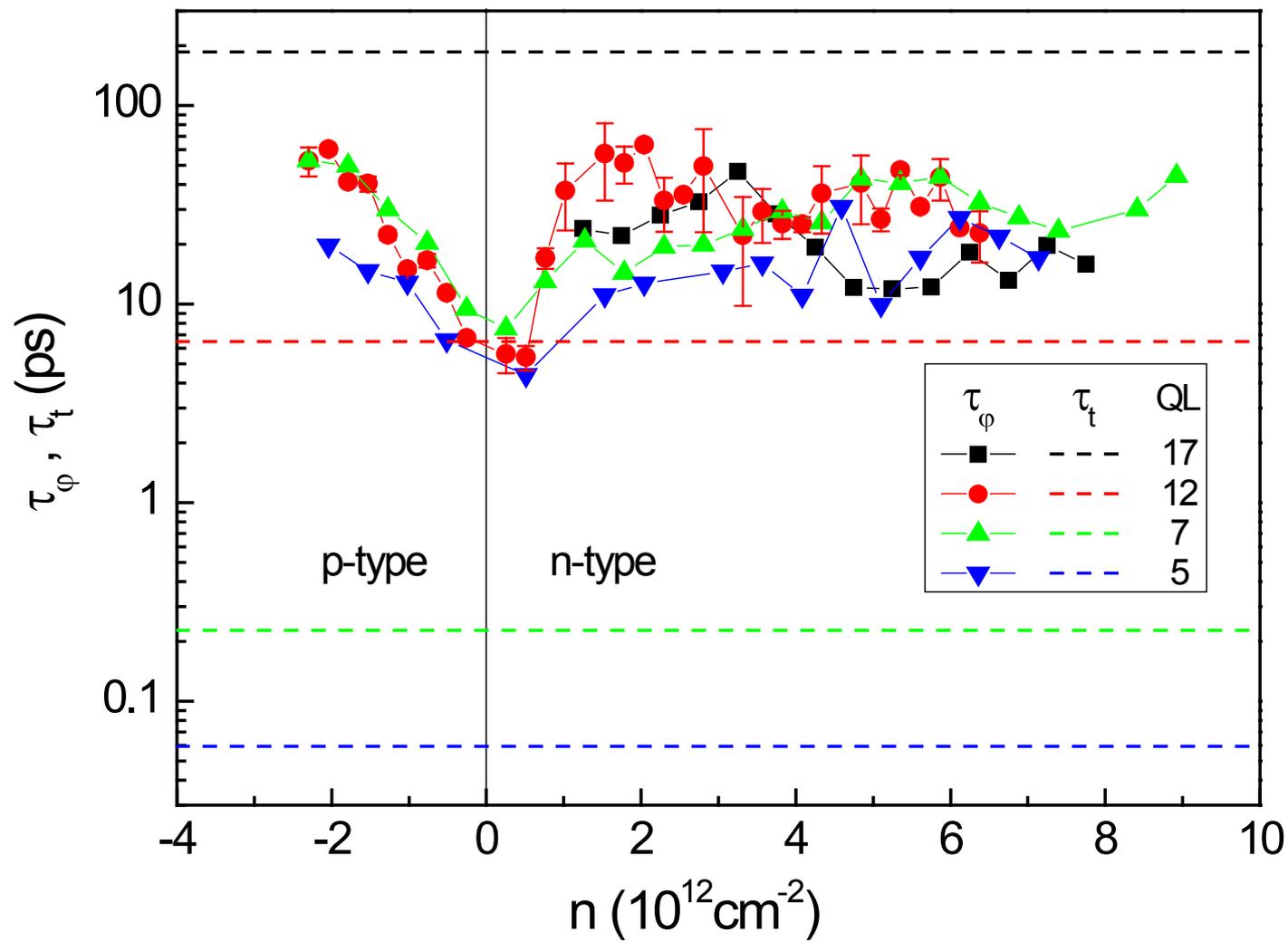

# Coherent Topological Transport on the Surface of Bi$_2$Se$_3$

Dohun Kim[1], Paul Syers[1], Nicholas P. Butch[2], Johnpierre Paglione[1], and Michael S. Fuhrer[1*]

*1. Center for Nanophysics and Advanced Materials, Department of Physics, University of Maryland, College Park, MD 20742-4111, USA*

*2. Condensed Matter and Materials Division, Lawrence Livermore National Laboratory, Livermore, CA 94550, U.S.A.*

**Supplementary Information**

1. **Charge transfer *p*-type doping with F4TCNQ : statistics and stability**

   Reducing unintentional doping in TIs is of central importance in realizing surface-dominant conduction. Here we adopt surface charge-transfer doping which provides simple and effective method for *p*-type doping TI thin films[5,11]. The doping agent we consider is 2,3,5,6-tetrafluoro-7,7,8,8-tetracyanoquinodimethane (F4TCNQ). It has high electron affinity (≈ 5.4 eV) and has been used as a *p*-type dopant in carbon nanotubes[1] and graphene[2]. We deposit ≈10 nm of F4TCNQ molecules on the Bi$_2$Se$_3$ Hall bar devices that are fabricated as described in the main text. Figure S1a shows Hall carrier density $n_H$ at zero gate voltage and at room temperature for 15 devices with (red) or without (blue) F4TCNQ deposition ranging in thickness from 5 to 17 nm. On average, the deposition of F4TCNQ results in a change of doping of -1 x 10$^{13}$cm$^{-2}$ (arrows in figure S1a); the negative sign indicates a shift toward *p*-type doping, i.e. a reduction in *n*-type doping.

Figure S1b shows conductivity $\sigma(n_H)$ of a F4TCNQ-doped 12 QL device at room temperature and ambient condition. The device were exposed to air for 1 (black square), 15 (red circle), and 25 (green triangle) days. The doping at zero gate voltage changes less than 10% in 15 days but shows negligible change from 15 to 25 days. We obtain the Hall mobility of ≈600 cm$^2$/Vs in all measurements, which shows good thermal stability of *p*-type doping by F4TCNQ molecules. This is in contrast to the behavior of Bi$_2$Se$_3$ device without F4TCNQ layer as surface degradation is commonly observed upon prolonged exposure to ambient[3,4]. Therefore, we conclude that F4TCNQ not only provides effective *p*-type doping but also act as stable capping layer for protecting the Bi$_2$Se$_3$ surface from degradation.

2. **Absence of activated insulating behavior in 5QL device.**

Here we discuss the possible origin of metallic conduction in gapped TI thin films. Figure S2a shows resistivity $\rho_{xx}$ of the 5 QL device as a function of gate voltage $V_g$ and temperature $T$. The shift of maximum resistivity $\rho_{max}$ (dashed line) is due to thermal activation from bulk valence band when the Fermi level $E_f$ is close to charge neutrality point[5]. As can be seen in figure S2b, $\rho_{xx}$ increases linearly with temperature for $T > 50K$, which is consistent with our previous study on electron-phonon scattering on the metallic TI surface[5]. The trace of $\rho_{max}(T)$ shows metallic behavior, i.e. $d\rho_{max}/dT > 0$ (inset of figure S2b). A weak insulating behavior ($d\rho_{max}/dT < 0$) can be seen for T < 40 K for data taken at fixed gate voltage (main panel figure S2b) but the temperature dependence is much weaker than thermal activation. Similar behavior was observed in MBE grown Bi$_2$Se$_3$ thin films and interpreted as electron-electron interaction corrections to the conductivity[6,7].

The hybridization induced energy gap Δ in few-QL $Bi_2Se_3$ has been measured in a recent ARPES study[8]. Transport experiment shows clearly insulating behavior in 3 QL thick $Bi_2Se_3$ field effect transistor, and energy gap as large as ≈250meV was estimated[9]. However, Δ is expected to decrease exponentially as a function of thickness. Moreover in the presence of finite disorder the surface, although gapped, can conduct through inhomogeneity-driven electron-hole puddles. For 5 QL device we estimate Δ from the fit to ARPES study[8] to be ≈34 meV. Previous scanning tunneling microscopy study of disorder driven electron-hole puddles in TIs[10] shows energy fluctuation of about 10~16 meV for doped $Bi_2Se_3$ with amount of disorder $n_{imp}$ ≈ $5 \times 10^{12} cm^{-2}$, which is comparable to Δ/2 ≈ 17 meV, and we generally estimate larger $n_{imp}$ in micro-fabricated devices[11]. This is a qualitative comparison only, since the screening will be poor for Fermi energies within the gap, and charge inhomogeneity will always create screening puddles with $E_F$ > Δ/2. However, the fact that the observed disorder scale for ungapped $Bi_2Se_3$ is comparable to the observed gap Δ is suggestive that our devices should be dominated by inhomogeneity. Therefore we conclude that the observed metallic temperature dependent resistivity in gapped TI thin films is due to conduction through inhomogeneity-driven electron-hole puddles.

Figure S1.

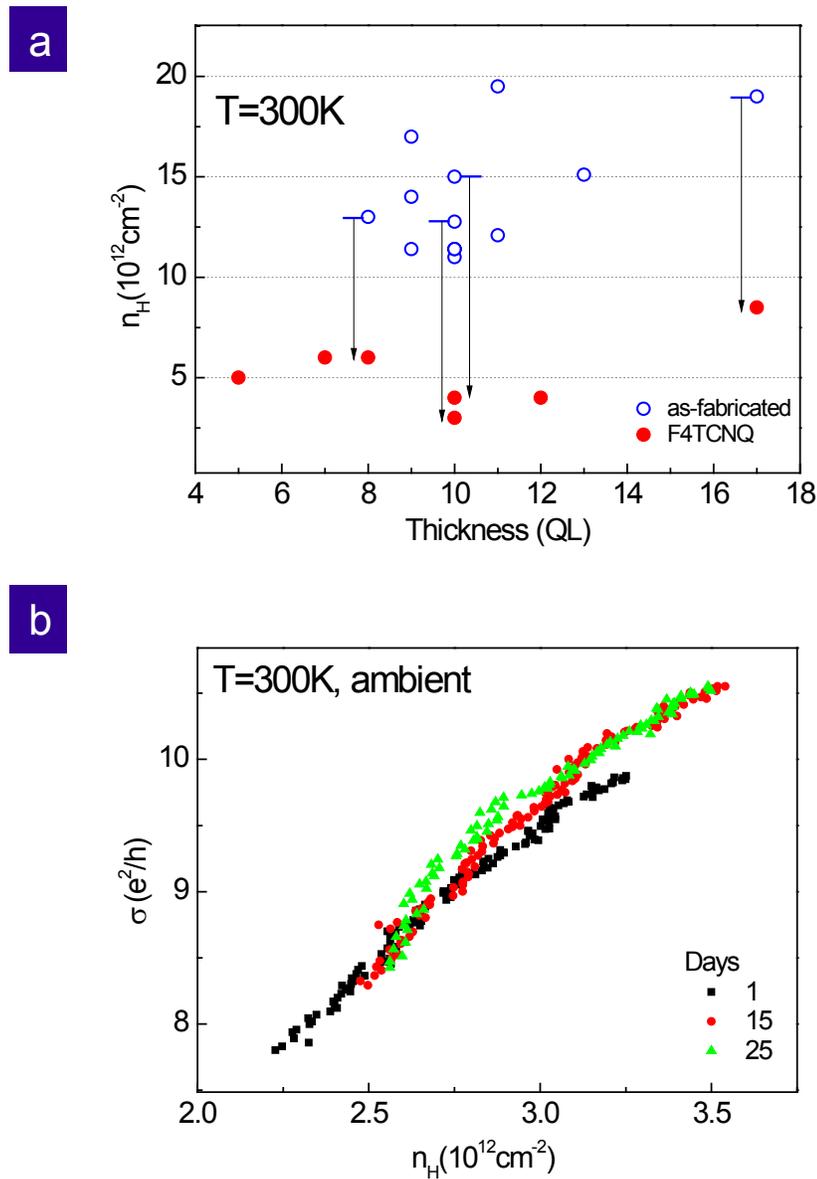

**Figure S1 Charge transfer *p*-type doping with F4TCNQ. a,** Hall carrier density $n_H$ at zero gate voltage and at room temperature for 15 devices with (red) or without (blue) F4TCNQ deposition ranging in thickness from 5 to 17 nm. Arrows indicate change of $n_H$ measured in the same sample. **b,** Conductivity σ of 12QL device as a function of $n_H$ at 300 K and ambient condition. The device were exposed to air for 1 (black square), 15 (red circle), and 25 (green triangle) days. In all measurements the range of gate voltage is from 0 to -20 V.

Figure S2.

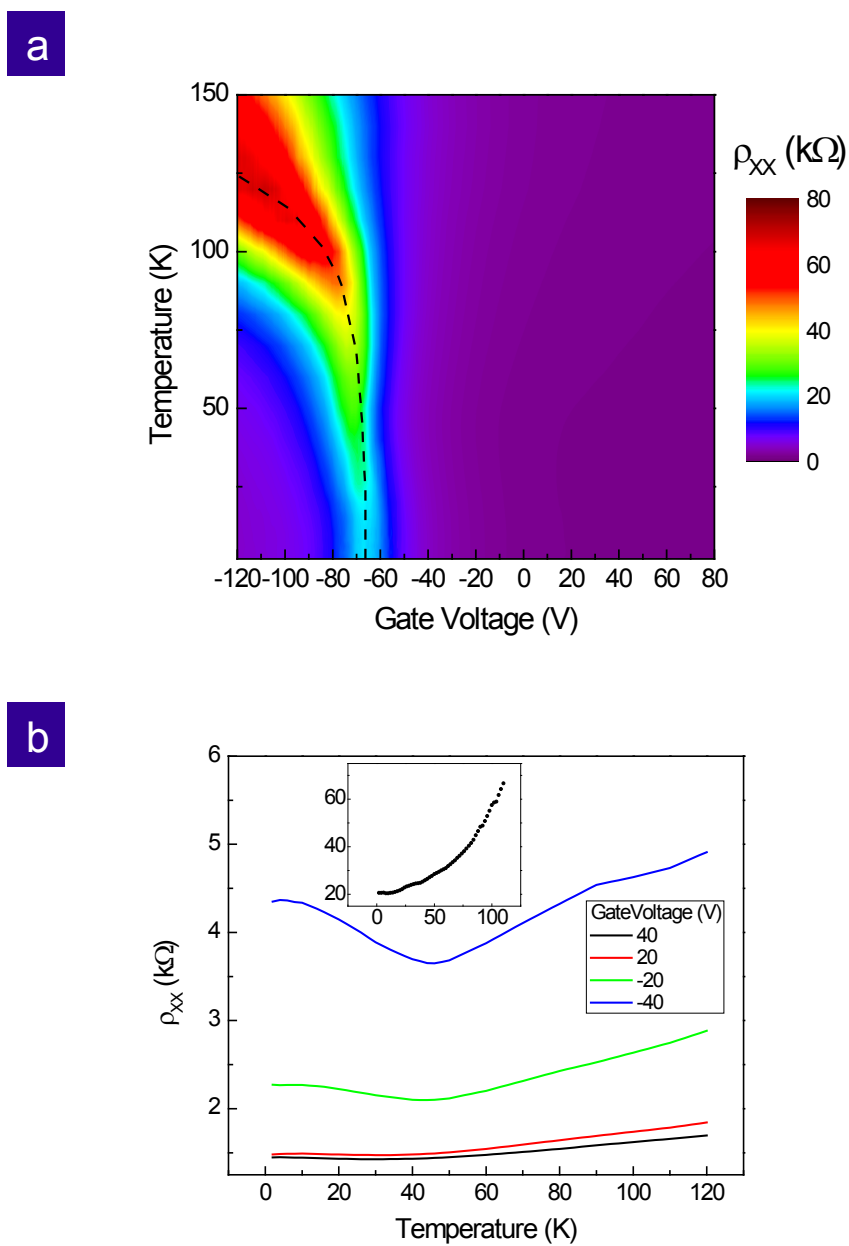

**Figure S2 Temperature dependent resistivity in 5QL device. a,** Longitudinal resistivity $\rho_{xx}$ of the 5 QL device as a function of gate voltage $V_g$ and temperature $T$. Dashed curve shows a trace of the position of $\rho_{max}$. **b,** $\rho_{xx}$ vs. $T$ at various gate voltages. The inset shows maximum resistivity $\rho_{max}$ as a function of $T$.